\documentclass[11pt]{article}
\usepackage{moriond,epsfig}

\begin{document}

\title{KAMLAND RESULTS}

\author{K.INOUE}

\address{Research Center for Neutrino Science, Tohoku University, \\
Aramaki Aoba, Aoba, Sendai, Miyagi 980-8578, JAPAN \\
E-mail: inoue@awa.tohoku.ac.jp}

\maketitle

\abstracts{
The LMA solution of the solar neutrino problem has been explored with 
the 1,000 ton liquid scintillator detector, KamLAND.
It utilizes nuclear power reactors distributing effectively $\sim$180 km
from the experimental site. Comparing observed neutrino rate with the 
calculation of reactor operation histories, an evidence for
reactor neutrino disappearance has been obtained from 162 ton$\cdot$year
exposure data. This deficit is only compatible with the LMA solution and 
the other solutions in the two neutrino oscillation hypothesis are excluded
at 99.95\% confidence level.
}

\section{Reactor Neutrinos}

In spite of the deficits observed in all solar neutrino experiments and 
improvement of their precisions, the solar neutrino problem has lasted 
for more than 30 years. A recent revolutionary neutral-current-measurement
by SNO experiment\cite{SNO} has provided an evidence of neutrino flavor transformation
and proved that the problem is really on neutrino characteristic. We also
know the evidence of atmospheric neutrino oscillation from SK\cite{SK}, and
it is natural to invest neutrino oscillation hypothesis to explain the solar neutrino
problem. In the two neutrino oscillation hypothesis, the large mixing angle 
(LMA) solution is the most preferable solution. However, another solution
also appears at 99\% confidence level and even completely different models such
as neutrino magnetic moment\cite{RSFP} and neutrino decay\cite{ND} can explain all the 
experimental data as well as the LMA solution.

\begin{figure}[th]
\begin{center}
\includegraphics[width=12cm]{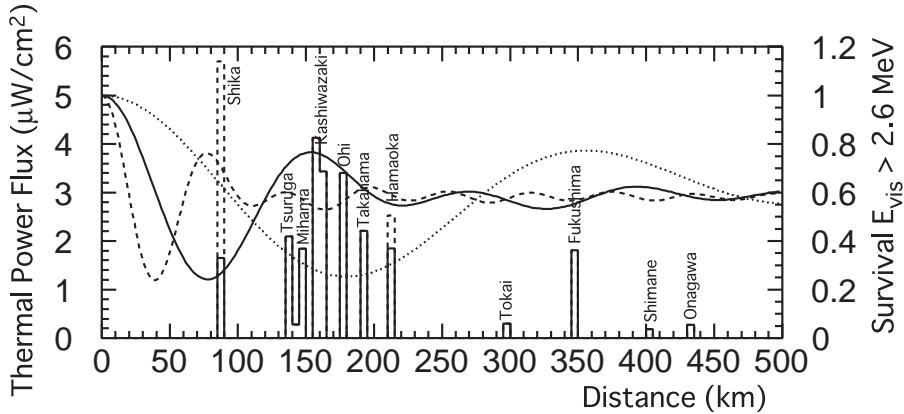}
\end{center}
\caption{Distribution of nuclear power reactors as a function of distance from
KamLAND site. Solid histogram is the current operation and dashed histogram is an
expected operation in 2006 (Shika at 88km increases by a factor 3).
Height of the histogram shows thermal power flux contribution at Kamioka.
Also shown as solid ($\Delta m^2=7\times10^{-5}$eV$^2$), 
dashes (3$\times 10^{-5}$) and dots ($1.4\times 10^{-4}$) lines are 
the survival probability of
$\bar{\nu}_e$ as a function of distance (all for $\sin^22\theta=0.84$). 
The probability is calculated for events above 2.6 MeV in visible energy.
\label{fig:distance}}
\end{figure}

It is necessary to use a well-understood artificial neutrino source to
meet a breakthrough in the solar neutrino study.
Anti-electron-neutrinos from nuclear power plants are such candidates.
Total power generation of world-wide reactors amounts to $\sim$1.1 TW 
and it corresponds to one third mole $\bar{\nu}_e$ creation per second.
Fortunate characteristic of KamLAND site is that
70 GW (7\% of world total) distributes at distances from 130 to 240 km
and consists 80\% of neutrino flux at Kamioka site, $\sim5\times10^{6}/cm^2/sec$.
This long baseline, effectively 180 km, is sensitive to the 
LMA oscillation parameters as shown in Fig.~\ref{fig:distance}.

Thermal power output of rector cores directly relates to the neutrino flux.
Those of all 54 Japanese commercial reactor cores 
are measured with strictly better than 2\% accuracy. In more detail, 
contributions from 4 different fissile nuclei,
$^{235}$U, $^{238}$U, $^{239}$Pu and $^{241}$Pu, have to be known for 1\% precision. 
Since the chemical composition of nuclear fuel changes in time as a burn-up effect,
we have to know the initial $^{235}$U enrichment and history of thermal power output
for each relevant reactor cores. This history information is also important to 
estimate time lag (0.28\% error) of beta decay from energy release 
and non-equilibrium effect\cite{noneq} of long-life nuclei (0.65\% error). 
We have obtained these information from all 54 Japanese cores.
The contribution of Korean reactors is 2.5\% and we obtained history of their electric power output.
Conversion of electric powers to thermal ones causes $\sim$10\% error
and, thus, Korean reactor makes 0.3\% error on neutrino flux estimation. 
All the other reactors contribute by only 0.7\% of total neutrino flux and error contribution
from them won't be larger than 0.35\%. Measured and/or calculated neutrino
spectra for 4 fissile nuclei\cite{spectra} make 2.48\% error on neutrino event rate.
Cross section of inverse beta decay is strongly related to free neutron life
time. Recent precise measurement of its life time (0.1\%) improved calculation of
the cross section\cite{cross} to better than 0.2\% with 1st order radiative corrections.

\section{KamLAND Detector}

KamLAND is the KAMioka Liquid-scintillator Anti-Neutrino Detector located
at the cavity for the former Kamiokande experiment.
It contains 1,200 m$^3$ liquid scintillator (LS: Pseudo-cumene 20\%, dodecane 80\% and
PPO 1.5g/l) and 1,800 m$^3$ buffer oil (BO) in 18 m diameter stainless steel tank.
Free protons in the LS are the $\bar{\nu}_e$ targets.
A positron from the inverse beta decay reaction ($\bar{\nu}_e + p \rightarrow e^+ + n$)
and a 2.2 MeV gamma ray from the neutron capture on a proton (mean capture time is $\sim$210 $\mu$sec)
make clear two-fold-delayed-coincidence signal.
The LS is suspended in the BO with an 135$\mu$m-thick-film balloon
(EVOH-Nylon-Nylon-Nylon-EVOH) and Kevlar rope network controled at neutral buoyancy (0.04\% heavier).
Photons emitted in the detector are monitored by 1325 newly developed 17" tubes and
554 old Kamiokande 20" tubes. While the total photo-coverage is 34\%, only the 17" tubes,
corresponding to 22\%, are used for this analysis. All these PMTs are isolated from
inner region by a 3-mm-thick acrylic sphere preventing radon emanation from these materials.
The outer detector (OD) is a water Cherenkov detector with 3.2 kton pure water and 225 old 
Kamiokande 20" tubes for purposes of external background absorption and cosmic-ray muon tag. 

Each photo-tubes are connected to two sets of three ranges 
analog-transient-waveform-digitizer recording whole pulse shape from one to thousands
of photo-electrons for about 200 nsec. Global triggers are issued based on number of
hit channels (each channels has about 0.3 p.e. threshold) and currently set at 200
hits ($\sim$ 700 keV) as a prompt trigger and at 120 hits ($\sim$400 keV) as delayed 
trigger for 1 msec after each prompt triggers. Current trigger rate is $\sim$25 Hz
and the data size amounts to 400 GB/day. This huge data size determines the lowest 
trigger threshold but it is sufficiently low for the reactor neutrino analysis
(all phenomena have visible energies more than two electron mass).
The OD trigger threshold is set to provide more than 99\% muon tagging efficiency.

The primary target, reactor neutrino measurements, 
requires radioactive impurity level of lower than 10$^{-13}$g/g for Uranium and
Thorium. These impurity level are measured by tagging $^{214,212}$Bi-$^{214,212}$Po 
in their decay chain and found to be $3.5\pm 0.5 \times 10^{-18}$ g/g of Uranium
and $5.2 \pm 0.8 \times 10^{-17}$ g/g of Thorium assuming radioactive equilibrium.
These values are much better than the initial requirement and even 
better than that of the future $^7$Be solar neutrino measurement. However, an extra purification
of $^{85}$Kr and $^{210}$Pb (daughter of $^{222}$Rn) 
contamination is necessary to start solar neutrino observation in the second stage.
Vigorous efforts are being made to invent an efficient purification method and
methods to measure their concentration.
Requirements and achievements are listed in Table.~\ref{table:purity}.
\begin{table}[th]
\caption{Requirements and Achievements of Radioactive Impurities. \label{table:purity}}
\vspace{0.4cm}
\begin{center}
\begin{tabular}{|c|r|r|r|r|}
\hline
{} &{} &{} &{} \\[-1.5ex]
Impurities & Achievements & Req.(reactor) & Req.(solar) \\[1ex]
\hline
$^{222}$Rn & 0.03 $\mu$Bq/m$^3$ & & \\[1ex]
$^{238}$U  & $3.5\pm 0.5 \times 10^{-18}$ g/g & $10^{-13}$ g/g & $10^{-16}$ g/g \\[1ex]
$^{232}$Th & $5.2\pm 0.8 \times 10^{-17}$ g/g & $10^{-13}$ g/g & $10^{-16}$ g/g \\[1ex]
$^{40}$K   & $<2.7 \times 10^{-16}$ g/g & $10^{-14}$ g/g & $10^{-18}$ g/g \\[1ex]
$^{85}$Kr  & $\sim $1 Bq/m$^3$ &  & 1 $\mu$Bq/m$^3$ \\[1ex]
$^{210}$Pb  & $\sim $100 mBq/m$^3$ &  & 1 $\mu$Bq/m$^3$ \\[1ex]
\hline
\hline
on the balloon &  & equiv. mine dust & \\[1ex]
\hline
$^{222}$Rn & $4.0 \times 10^{-4}$ Bq & & \\[1ex]
$^{238}$U  & $3.1 \times 10^{-8}$ g & 0.9 g &  \\[1ex]
$^{232}$Th & $9.7 \times 10^{-4}$ g & 0.1 g &  \\[1ex]
\hline
\end{tabular}
\end{center}
\end{table}

\section{Calibrations and Systematic errors}

Energy calibrations are performed suspending radioactive sources 
($^{68}$Ge, $^{65}$Zn, $^{60}$Co and AmBe) along
the z-axis. They covers energy range from 0.5 MeV to 7.6 MeV.
Spallation products (neutron, $^{12}$B/$^{12}$N and $^8$He/$^9$Li) are also utilized 
to know the behavior at off z-axis and in higher energies up to $\sim$15 MeV.
They distribute uniformly both in time and space and thus very useful to monitor
space uniformity and time variation. 
Gamma rays ($^{40}$K and $^{208}$Tl) from external material also 
provided a good monitor of time variations.
In addition to these wide energy range calibrations, alpha decays in Bi-Po chain provides wide variety
of dE/dx and it helped detail study of quenching effect of LS (thus linearity study
of energy scale).
Observed uniformity of energy scale is better than 0.5\% in 5-m-radius fiducial volume
and time variation of the scale was controlled within 0.6\%.
The systematic error for the 2.6 MeV energy threshold is estimated as 2.13\% in 
neutrino event rate.
Observed photon yield is $\sim$300 p.e./MeV at the center only with 17" PMTs and it
will become about 500 p.e./MeV when we start to use 20" PMTs. Current energy resolution
is $\sim 7.5\%/\sqrt{E}$.

\begin{figure}[th]
\begin{center}
\includegraphics[width=10cm]{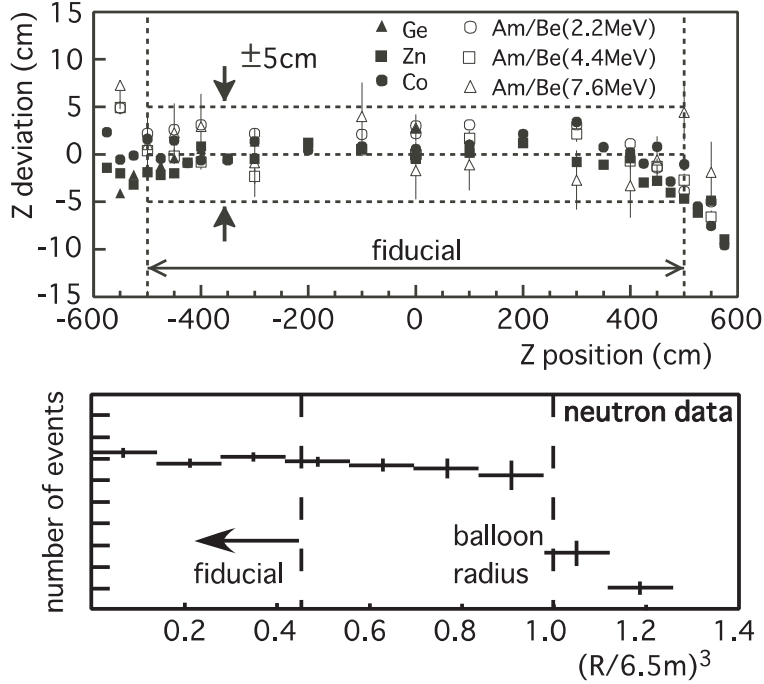}
\end{center}
\caption{Vertex calibrations. Upper panel shows systematic bias of
reconstructed positions along z-axis. Biases are less than 5 cm within fiducial
region -500 cm to 500 cm for all energies. Lower panel is the R$^3$ distribution 
of uniformly distributing spallation neutron events. 
Uniform distribution appears as flat distribution.
\label{fig:vertex}}
\end{figure}

The LS density is 0.780 g/cm$^3$ at 11.5 degree and, thus, the number of free protons 
in the 5-m-radius fiducial volume is $3.46\times 10^{31}$. 
The fiducial cut is applied based
on the reconstructed vertices from the relative times of PMT hits (typical
resolution is $\sim$25 cm for 2 MeV events).
As shown in Fig.~\ref{fig:vertex}, observed systematic biases along z-axis are smaller than 
5 cm in the entire fiducial range.
It corresponds to less than 3\% fiducial volume error under spherical symmetry 
of the detector. This systematic error was verified with uniformly distributing spallation
events (gamma ray from neutron capture and $^{12}$B beta decays). These spallation
events covers energy range from 2.2 MeV to $\sim$15MeV where all relevant energies
to the reactor neutrino analysis is included. Event rates in the fiducial and in the total
volume are compared with the volume ratio and their precision 
is considered as the systematic error of the fiducial volume.
Neutron data gave $-1.48\pm2.58\%$ and $^{12}$B data gave $+0.16\pm3.34\%$.
This verification is currently limited by statistics of spallation events.
We employ the most conservative value $1.48\% + 2.58\% \rightarrow \pm4.1\%$ 
and accounting for uncertainty in the LS total mass of 2.13\%, we assign 4.6\%
as the systematic error of the number of target protons.

Detection and tagging efficiencies of delayed coincidence signals are
measured by a LED pulser, intensity measurements of radioactive sources,
AmBe delayed coincidence signals and so on. For the specific selection
criteria applied in this analysis; (1) a 5-m-radius fiducial cut,
(2) a time correlation cut (0.5 $\mu$sec to 660 $\mu$sec), (3) a vertex correlation
cut (1.6 m), (4) a delayed energy cut (1.8 MeV to 2.6 MeV), and (5) a cylindrical cut
around z-axis for delayed signal (1.2 m), we obtained $78.3\pm 1.6$\% efficiency
for the reactor neutrino signals.

Total systematic error we estimated is 6.42\% for 2.6 MeV analysis threshold.
And break down of the error is listed in Table.~\ref{table:errors}.
\begin{table}[th]
\caption{Estimated systematic uncertainties in \%. \label{table:errors}}
\vspace{0.4cm}
\begin{center}
\begin{tabular}{|ll|ll|}
\hline
{} &{} & {} & {}\\[-1.5ex]
Thermal power output (Japanese) & 2.0 & Cross section      & 0.2  \\[1ex]
Korean reactors                 & 0.3 & Total LS mass      & 2.13 \\[1ex]
Other reactors                  & 0.35& Fiducial ratio     & 4.06 \\[1ex]
Burn up effect                  & 1.0 & Energy threshold   & 2.13 \\[1ex]
Long-life nuclei                & 0.65& Efficiency of Cuts & 2.06 \\[1ex]
Time lag of beta decay          & 0.28& Live time          & 0.07 \\[1ex]
Neutrino spectra                & 2.48&                    &      \\[1ex]
\hline
Total & & & 6.42\% \\[1ex]
\hline
\end{tabular}
\end{center}
\end{table}

\section{Backgrounds}
The most difficult background in reactor neutrino analysis is geo-neutrinos. 
The earth has 44 TW heat flow at the surface. Twenty TW of it
is thought to come from radio-activities in the earth, 16 TW from
Uranium and Thorium and 4 TW from $^{40}$K. In U and Th decay
chains, there are beta decays emitting observable $\bar{\nu}_e$ 
with the inverse decay (1.806 MeV threshold). They have identical signature
with reactor neutrinos. Thier sharp edges at the end-points 
will be useful to distinguish them when statistics is sufficient. 
We currently cannot subtract their contributions blindly using inaccurate predictions. 
Then, the analysis threshold of visible energy, 2.6 MeV,
is set above the geo-neutrino end-point energy, 2.49 MeV.
On the other hand, we can separately obtain U and Th contributions 
from their characteristic energy spectra when sufficient statistics is acquired.
Subtraction of nearby contributions with a radio-activity map will make possible to 
investigate interior of the earth with neutrinos. This is the start of the new field
"Neutrino Geophysics."
The lower threshold covering all $\bar{\nu}_e$ events (0.9 MeV) is also 
used for a consistency check and geo-neutrino search.

Accidental backgrounds are expected to be only $1.81\pm0.08$ and $0.0085\pm0.0005$
events in final data sets of 0.9 and 2.6 MeV thresholds, respectively.
Main fakes are from $^{210}$Bi as prompt and $^{208}$Tl as delayed events.

We also have two types of correlated backgrounds associated by energetic cosmic-ray muons.
One is fast neutrons from outside and another is long-lived beta-decay-nuclei associating
neutron emission in the detector.
Contribution of fast neutrons are estimated by tagging muons in the OD and look for
delayed coincidence signal in the ID. Vertices of such events are concentrated close
to the wall, and there are no events entering the fiducial volume. We obtained upper
limit of such events. Considering inefficiency of the OD tagging and estimating 
contribution of rock penetrating muons by simulation with restriction of this measurement,
we obtained fast neutron backgrounds be smaller than 0.5 events.

Long-lived neutron emitters are $^8$He and $^9$Li. Their half-lives are 0.12 and 0.18 sec
and 16\% and 50\% of their beta decays they also emit neutrons, respectively.
In order to eliminate these backgrounds, spallation cuts are employed.
We apply 2 msec veto after any muons.
For muons with extra energy losses from minimum ionization larger than $10^6$ p.e. ($\sim$3 GeV),
two seconds veto is additionally applied. For smaller energy losses, 2 sec veto is
applied only in 3-m-radius cylinder around muon track. These spallation cuts cause 11.4\% 
dead time. After these cuts, we expect $1.1\pm1.0$ and $0.94\pm0.85$ events for two analysis
thresholds.

Summary of backgrounds is shown in Table.~\ref{table:backgrounds} together with 
a model prediction of geo-neutrino events.
\begin{table}[ht]
\caption{Summary of backgrounds (events/dataset). \label{table:backgrounds}}
\vspace{0.4cm}
\begin{center}
\begin{tabular}{|l|c|c|}
\hline
{} &{} & {} \\[-1.5ex]
Backgrounds   & 0.9 MeV threshold & 2.6 MeV threshold \\[1ex]
Accidental    & $1.81\pm0.08$     & $0.0085\pm0.0005$ \\[1ex]
$^8$He/$^9$Li & $1.1\pm1.0$       & $0.94\pm0.85$     \\[1ex]
Fast neutron  & $<0.5$            & $<0.5$            \\[1ex]
\hline
Total         & $2.91\pm1.12$     & $0.95\pm0.99$     \\[1ex]
\hline
Geo$\bar{\nu}_e$ 16 TW & 9.1  & 0.044     \\[1ex]
\hline
\end{tabular}
\end{center}
\end{table}

\begin{figure}[th]
\begin{center}
\includegraphics[width=11cm]{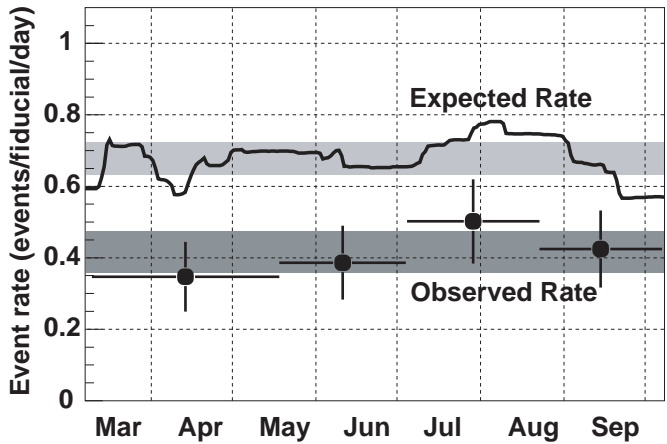}
\end{center}
\caption{Reactor neutrino event rate.
Plots and line are the observed and expected event rate, respectively, and
gray hatches are their averages. Structure in expected rate reflects change of reactor
operations.  \label{fig:rate}}
\end{figure}

\section{Results}
The data used for the analysis is from March 4th to October 6th, 2002.
Total live time is 145.1 days (after dead time subtraction) and it corresponds to 162 
ton-year exposure.
As seen in Fig.~\ref{fig:rate}, observed rates are always smaller than the expected ones
from no oscillation. While expected number of neutrino events above 2.6 MeV 
is $86.8\pm 5.6$, we observed only 54 events including $0.95\pm 0.99$ background
events resulting neutrino disappesarance at 99.95\% confidence level.
The ratio (Observed-B.G.)/no-oscillation is $0.611\pm0.085(stat.)\pm0.041(syst.)$.
In the 0.9 MeV threshold dataset, we expect $124.8\pm 7.5$ reactor neutrino events,
$2.91\pm 1.12$ backgrounds and also $\sim$9 geo-neutrino events from a model.
However, we observed only 86 events and low energy data is also consistent with the deficit
above 2.6 MeV.

\begin{figure}[th]
\begin{center}
\includegraphics[width=10cm]{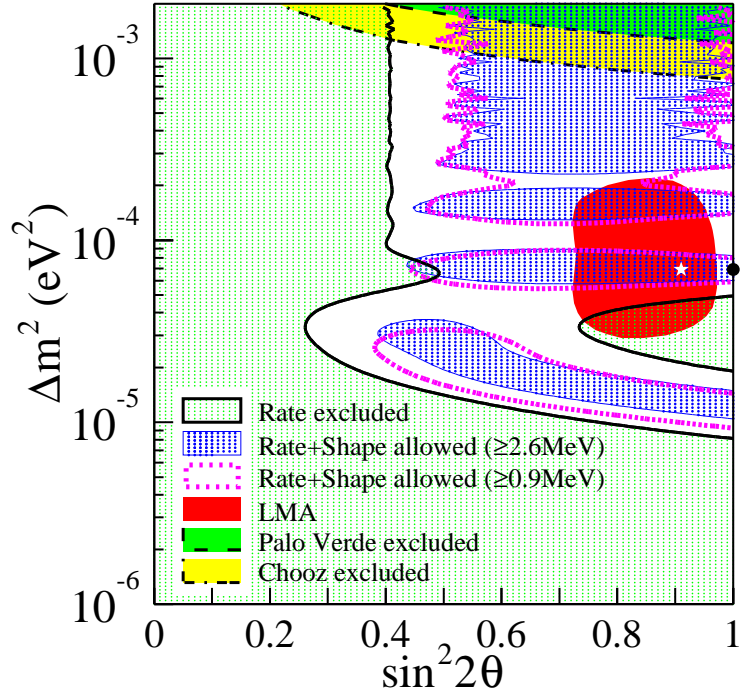}
\end{center}
\caption{Excluded region by rate analysis and allowed region by rate+shape analysis
are shown. All region are shown for 95\% confidence level. The LMA is obtained
from combined analysis of solar neutrino observations.
\label{fig:contour}}
\end{figure}

This evidence of neutrino disappearance supports the LMA solution of the solar neutrino
problem and all the other oscillation solutions are excluded at 99.95\% C.L. 
under the CPT invariance. Also the other exotic models (RSFP, neutrino decay etc) can
not be the leading phenomena of the solar neutrino problem. Adding the KamLAND results
to the solar neutrino observations, we finally solved the solar neutrino problem and the
LMA solution is the right solution.
\begin{figure}[th]
\begin{center}
\includegraphics[width=10cm]{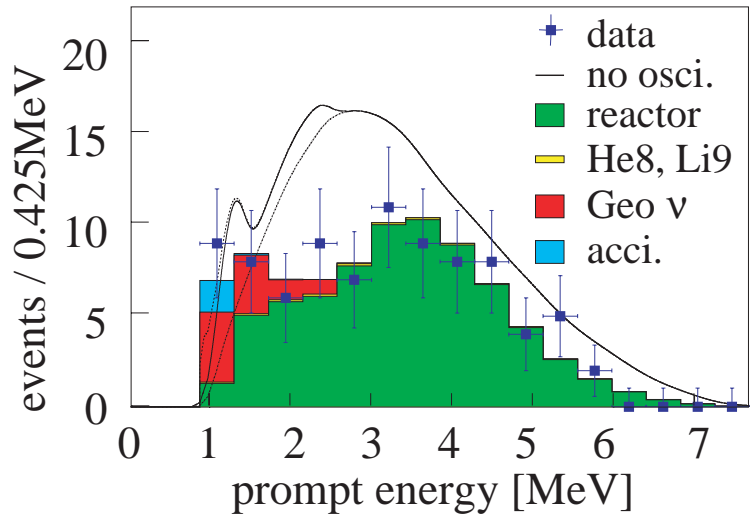}
\end{center}
\caption{Observed energy spectrum is shown together with the expected 
from no oscillation and best fit oscillation spectra. Best fit backgrounds
are also shown.
\label{fig:spectrum}}
\end{figure}
In order to claim the neutrino oscillation with KamLAND data alone, spectrum distortion
has to be observed. Currently, the distortion (see Fig.~\ref{fig:spectrum}) is not 
significant to claim it but
helps to shrink the allowed region of the oscillation parameters.
Fig.~\ref{fig:contour} shows the excluded region from the rate analysis and 
allowed region from rate + shape analysis with two different threshold data.
Only two bands overlap with the LMA region\cite{global} from solar observations.
Two different thresholds give similar allowed region and it means
data below and above 2.6 MeV are consistent with each other. When using data below
2.6 MeV, both Uranium and Thorium geo-neutrino contributions are treated as free parameters.
This is the reason that the allowed region doesn't shrink with larger statistics in 0.9 MeV
threshold. The best fit parameters are met at 
($\sin^22\theta$,$\Delta m^2$)=(1.0,$6.9\times10^{-5}$eV$^2$) with 2.6 MeV threshold and
(0.91,$6.9\times 10^{-5}$) with 0.9 MeV threshold. The mass difference is stable
but mixing angle changes easily. In the low threshold analysis, we obtained the best fit
value for U and Th as 4 and 5 events. It corresponds to $\sim$40 TW heat
flow from Uranium and Thorium. But 0 to 110 TW is allowed at 95\% C.L.
and statistics is yet insufficient to claim an observation of geo-neutrinos.

KamLAND results shrank the allowed region of oscillation parameters to two bands,
LMA1 ($\Delta m^2\sim 7\times 10^{-5}$ eV$^2$) and LMA2
($\Delta m^2\sim 1.4\times 10^{-4}$ eV$^2$). Chi-square of LMA2 is about 3
worse than that of LMA1 with current statistics. If LMA1 is the true answer, 
the LMA2 may be excluded in the next update in a year. 
In any cases, new Shika reactor starts in 2006 locates at a good distance to 
discriminate them. As shown in Fig.~\ref{fig:distance}, 
Shika distance is close to the first minimum of the LMA1
and the LMA2 has the second maximum at the distance.
Comparison before and after 2006 will give good separation of these two solutions.

Our current efforts are focused on the purification of the LS aiming at
observing $^7$Be solar neutrinos with neutrino-electron scattering. 
It will make a next milestone in the "Neutrino Astrophysics" and also by combining 
with charged current observations, it will give a precise measurement of mixing angle.

\section{Summary}
KamLAND observed an evidence of reactor neutrino disappearance at $\sim$180 km baseline.
Only the LMA solution 
is compatible with this disappearance and the long standing solar neutrino problem
is finally solved with a combination of KamLAND and solar neutrino
observations. KamLAND will give a precise value of $\Delta m^2$ and 
the mixing angle with reactor and solar $^7$Be solar neutrino observations.
Current hint of the geo-neutrino flux will soon become first
observation of geo-neutrinos.

This manuscript is based on the paper\cite{paper} and the KamLAND experiment
is successfully operated and performed by efforts of the KamLAND collaboration from
Japan, United States and China.

\end{document}